\documentclass[aps,10pt,prd,twocolumn,superscriptaddress,preprintnumbers,showkeys,nofootinbib,showpacs]{revtex4-1}

\usepackage{bm}
\usepackage{graphicx}
\usepackage[hypertexnames=false, colorlinks=true, citecolor=blue, urlcolor=blue, linkcolor=black]{hyperref}
\usepackage{amsmath,amssymb,amsfonts,latexsym,empheq,float}
\usepackage{wasysym}
\usepackage{refcount}
\usepackage[normalem]{ulem} 
\usepackage{natbib}
\usepackage{hypernat}
\usepackage{float}
\usepackage[amsmath]{ntheorem}
\begin{document}
\author{J.~M.~Alarc\'on}
\email[ E-mail: ]{jmanuel.alarcon@uah.es}
\affiliation{Universidad de Alcal\'a, Grupo de F\'isica Nuclear y de Part\'iculas, Departamento de F\'isica y
Matem\'aticas, 28805 Alcal\'a de Henares (Madrid), Spain}
\author{E. Lope-Oter}
\email[ E-mail ]{mariaevl@ucm.es}
\affiliation{Departamento de F\'{\i}sica Te\'orica \& IPARCOS,\\ Universidad Complutense de Madrid, E-28040 Madrid, Spain}
\author{J. A. Oller}
\email[ E-mail: ]{oller@um.es}
\affiliation{Departamento de Física, Universidad de Murcia, E-30071 Murcia, Spain}
\preprint{IPARCOS-UCM-24-049}

\title{Regulator-independent equations of state for neutron stars generated from first principles}

\begin{abstract}
We study the equation of state (EoS)  of a neutron star (NS) accounting for new advances. In the low energy density,  $n\leq 0.1 n_s$, with $n_s$ the saturation density, we use a new pure neutron matter EoS that is regulator independent and expressed directly in terms of experimental nucleon-nucleon scattering data. In the highest-density domain our EoS's are matched with pQCD to ${\cal O}(\alpha_s^3)$. First principles of  causality, thermodynamic consistency and stability are invoked to transit between these two extreme density regimes. The EoS's are further constrained by the new measurements from PREX-II and CREX on the symmetry energy ($S_0$) and its slope ($L$). In addition, we also take into consideration the recent experimental measurements of masses and radii of different NSs and tidal deformabilities.  A band of allowed EoS's is then obtained. Interestingly, the resulting values within the band for $S_0$ and $L$ are restricted with remarkably narrower intervals than the input values, with  $32.9\leq S_0 \leq 39.5~\text{MeV}$ and $   37.3 \leq L\leq 69.0~\text{MeV}$ at the 68\% CL.  The band of EoS's constructed also allows possible 
 phase transitions (PTs) for NS masses above 2.1~$M_\odot$ at 68\% CL for $n>2.5n_s$. We find both long and short coexistence regions during the PT, corresponding to first and second order PTs, respectively.  We also generate the band of EoS's when excluding the astrophysical observables. 
This is of interest  to test General Relativity and modified theories of gravity. Our band of EoS's for NSs can be also used to study other NS properties and dark matter capture in NS. 

\end{abstract}

\maketitle

\section{Introduction}
 
 Neutron stars (NSs) are among the most dense objects in the universe, and are formed from supernovae explosions of massive stars.
 The extreme conditions in these compact objects provide a way to study the nuclear matter in a wide range of densities \cite{Lovato:2022vgq}.
 Neutron stars have been also proposed to study physics beyond the standard model, as dark matter detection, since the intense gravitational field can enhance the capture rate of dark matter particles \cite{Bell:2020jou}.  
In addition, properties of  NSs as the tidal deformability, are directly related to the properties of the gravitational-wave signals from a binary NS system \cite{Raithel:2019uzi}.
The equation of state (EoS) of a  NS is one of the needed inputs for these studies, and its calculation requires techniques employed in nuclear physics and in QCD. 

A NS is typically divided into three main regions: the atmosphere, the crust (outer and inner), and the core (outer and inner). The central number  density of a NS can reach values that are typically between 5 to 10 times higher than the saturation density.
The crust is composed of inhomogeneous nucleonic matter and unlocalized $e^-$ in $\beta$-equilibrium \cite{Baym:1971ax,Negele:1971vb,Sharma:2015bna}.   
The internal constitution of the outermost layers of the crust is  well known and the equation of state of this region is determined by experimental atomic mass measurements.  
The inner crust of a NS cannot be reproduced in terrestrial laboratories and, therefore, the description of the inner crust must rely on theoretical models. Moving towards the center of the NS, the number density increases, and the nuclei in the crust progressively become more  neutron-rich. Eventually, at a certain point known as the neutron-drip point, the number density becomes high enough that neutrons start to leak out from the nuclei, and the inner crust begins. In this regard, at about half the saturation density $n=n_s/2 = 0.08 $ fm$^3$, where $n$ is  the number baryon density and $n_s=0.16$~fm$^{-3}$ is the saturation density, corresponding to an energy density $\varepsilon_s \approx 1.8\times 10^{14}$~g/cm$^3=70$~MeV fm$^{-3}$, the nuclei completely melts and the NS core would begin. This region is much less understood than the crust and it is divided into outer and inner parts, as in the crust. The core of a NS is  supposedly composed of uniform matter at  beta-equilibrium, whose properties are  strongly related to the symmetry energy originating from the energy difference between pure neutron matter (PNM) and symmetric nuclear matter (SNM).

Direct experimental constraints on NS EoS has improved drastically during recent years thanks to measurements of masses by Saphiro time delay \cite{Demorest:2010bx,Antoniadis:2013pzd,Fonseca:2021wxt}, NICER telescope of  masses and radii from X-ray data \cite{Riley:2019yda,Miller:2019cac,Riley:2021pdl,Miller:2021qha,Salmi:2022cgy}, and gravitational-wave detection from binary NS mergers by the LIGO and Virgo Collaborations \cite{Abbott:2018wiz,Abbott:2018exr,LIGOScientific:2020aai}. Needless to say that all of these data have spurred many theoretical studies to infer valid EoS's \cite{Koehn:2024set,Annala:2019puf,Raaijmakers:2021uju,Pang:2021jta,Legred:2021hdx,Biswas:2021yge,Biswas:2021paf,Ecker:2022dlg,Altiparmak:2022bke,Huth:2021bsp,Annala:2023cwx,Somasundaram:2021clp,Essick:2023fso,Lim:2022fap,Brandes:2023hma,Brandes:2023bob,Tang:2023owf,Llanes-Estrada:2019wmz}. In this work, we also attend to this new vigorous impetus on inferring the EoS out of the NS properties. Our novelty here is that we incorporate an EoS for pure PNM at zero temperature and very low densities directly expressed in terms of the nucleon scattering data \cite{Alarcon:2022vtn}, i.e. phase shifts and mixing angles. Let us stress that this EoS is a renormalized result without any regulator dependence, like a cutoff. It stems from the calculation of the neutron-neutron scattering amplitude  in PNM as a sum of partial-wave amplitudes, which is then used to resum the ladder series \cite{Thouless:1960ann,Alarcon:2021kpx,Alarcon:2021nwd,Alarcon:2022vtn} for obtaining the energy per neutron $E/A$. 
This EoS at low number density is interpolated up to known higher-density results from perturbative QCD (pQCD), applying causality, monotonicity and thermal consistency \cite{Komoltsev:2021jzg}, in two steps. We use a first interpolation between the uncertainty band resulting for the aforementioned EoS \cite{Alarcon:2022vtn} for PNM  and  saturation density, constrained  by nuclear experimental data regarding the values of the symmetry energy and its slope \cite{PREX:2021umo,CREX:2022kgg}. Then, we use a second interpolation between this first band and the high-density pQCD regime \cite{Kurkela:2009gj,Fraga:2013qra,Gorda:2021kme}, so we cover a whole range of  baryon densities up to around $7~$fm$^{-3}$.  
All PTs that we find occur for $M\geq 2.1\,M_\odot$ at 68 $\%$ CL  within the range of starting densities between $2.51 n_s \leq n \leq n_{c}$.

In this last step, we distinguish between two types of procedures, depending on whether we take into account astrophysical observables or not. In one case we use General Relativity (GR) theory to calculate mass and radii of NSs, while in the other only terrestrial experimental results from nuclear physics are accounted for. This is interesting  because one can then use  the resulting EoS's in the latter case to test modified gravity theories \cite{LopeOter:2019pcq,Lope-Oter:2021vxl}. 
 
The contents of the manuscript are organized as follows. 
In Sec. \ref{method} we present the method use to compute the EoS at different baryon density regions. In Sec. III we show the results obtained for the different regions, when considering or not astrophysical observables for the computation of the EoS. The results are compared to measurements of masses and radii of different neutron stars, and the tidal deformability-mass relation of GW170817. A summary of the work and conclusions are explained in Sec. IV.

\section{Methodology}
\label{method}

The number density inside a neutron star can reach several times the saturation density. 
Therefore, the calculation of the equation of state combines results in the low and high density regime. 
In the low density regime,  approaches used in nuclear physics as effective field theories (EFTs) or meson exchange models have been typically employed. 
In the high-density domain, the pQCD results are normally used  to constrain the EoS at densities higher than the saturation one. Calculations in pQCD are now available up to ${\cal O}(\alpha_s^3)$ \cite{Kurkela:2009gj,Gorda:2021znl,Gorda:2021kme}. This information can be  extrapolated down to lower densities \cite{Komoltsev:2021jzg} as we apply below. 
One can also use experimental information at the nuclear saturation density to constrain further the EoS at intermediate densities.
In the following, we explain the strategies used in the different density regions to  obtain the EoS of pure neutron matter that will be applied to calculate properties of the neutron stars.


\subsection{Low density region}
For the low number density region we use the result for the energy per nucleon ($E/A$) obtained in \cite{Alarcon:2022vtn} by two of the authors.
This calculation differs from the standard EFT calculations in several aspects. First, it is a renormalized calculation, and the results for the energy per particle ($E/A$) are independent of any regulator. However, in present EFT calculations based on Chiral Perturbation Theory  there is an explicit and significant (power like) dependence in an unphysical cutoff, such that typically very small variations in this parameter are allowed when discussing results \cite{Drischler:2020yad,Drischler:2017wtt}. 
Second, the results for $E/A$ obtained \cite{Alarcon:2022vtn} are expressed in terms of experimental scattering data from nucleon-nucleon interactions in vacuum, namely, phase shifts and mixing angles \cite{NavarroPerez:2014ovp}. However, in EFT calculations a potential has to be first worked out. This implies that its many counterterms needed for a higher-order calculation, like the N$^3$LO in Ref.~\cite{Drischler:2017wtt,Drischler:2020yad}, necessarily have to be fitted to a diversity of data, with the caveats that so-many free-parameter fits always arise. Third, the chiral EFT calculation in the nuclear medium naively apply the same power counting as in vacuum, while it is well-known \cite{Meissner:2001gz,Oller:2009zt,Oller:2019ssq} that the counting gets modified due to the presence of infrared enhanced nucleon propagators when the energy (minus the nucleon mass in a relativistic treatment) through the propagator is ${\cal O}(k_F^2/m)$. This implies that the leading order calculation of  $E/A$ in the nuclear medium requires to resum the series of ladder diagrams \cite{Lacour:2009ej,Alarcon:2021kpx,Alarcon:2022vtn}. This is a non-perturbative calculation beyond standard chiral EFT calculations. The formalism for performing this resummation was developed in Ref.~\cite{Alarcon:2021kpx}, and then applied to nuclear matter in Ref.~\cite{Alarcon:2022vtn}. This theory requires the non-perturbative calculation of the in-medium nucleon-nucleon scattering amplitude  by solving the corresponding integral equation \cite{Alarcon:2021kpx}.  Indeed, as explained in Ref.~\cite{Alarcon:2022vtn}, and also recognized in Ref.~\cite{Hebeler:2010xb} within the realm of chiral EFT in-medium calculations, in the dilute limit for nuclear matter non-perturbative effects are important  because the nucleon-nucleon system is clearly non-perturbative there due to of the presence of the Deuteron in the isoscalar channel and the antibound state in the isovector one, the case of PNM.\footnote{This is the reason why higher-order chiral EFT calculations of reference, like those in \cite{Drischler:2020yad,Drischler:2017wtt}, show $E/A$ for $n>0.05~$fm$^{-3}\approx 0.3 n_s$.} The integral equation to calculate the nucleon-nucleon partial-wave amplitude in nuclear matter was solved  in the limit of contact interactions, and this is why the resulting EoS from Ref.~\cite{Alarcon:2022vtn} is valid for $k_F\lesssim 150$~MeV, or up to a density of $n\lesssim 0.015~$fm$^{-3}\approx 0.1~n_s.$ 
An error estimated for having taking only contact interactions is also evaluated in Ref.~\cite{Alarcon:2022vtn} by including a Gau{\ss}ian  prefactor in momentum space controlled by a scale $Q$, with a value around the pion mass. Estimates for higher order beyond the ladder-resummation leading-order result for $E/A$ were also considered, e.g. by appropriately employing a density-dependent neutron mass taken from Ref.~\cite{Huth:2020ozf}. These latter effects were typically much smaller than the uncertainty due to the Gau{\ss}ian cutoff smearing function (see  
Ref.~\cite{Alarcon:2021kpx,Alarcon:2022vtn} for the full formalism and technical details).

The upper value $n\approx 0.015~$fm$^{-3}\approx 0.1 n_s$ for direct applicability of the EoS \cite{Alarcon:2022vtn}, we call this density limit $n_L$, is in the  density range of the crust of \cite{Baym:1971ax, Sharma:2015bna}. Thus, no use of any crust EoS is implemented in our interpolations. However, by comparing the calculation  in Ref.~\cite{Sharma:2015bna} for the difference in $E/A$ between  uniform nucleon matter in $\beta$ equilibrium and the actual most favorable energetic configuration for the inner crust we observe that this difference at $0.015$~fm$^{-3}$ amounts only  
to around $-0.4$~MeV, which is well inside the uncertainty band estimated in Ref.~\cite{Alarcon:2022vtn} at this density. This is clear by comparing Fig.~3 in Ref.~\cite{Sharma:2015bna} with Fig.~5 of Ref.~\cite{Alarcon:2022vtn}. Additionally, figure~5 of Ref.~\cite{Sharma:2015bna} also predicts a proton fraction of only $\sim 3\%$ at this density. 

\begin{figure}
     \centering
\includegraphics[width=0.39\textwidth]{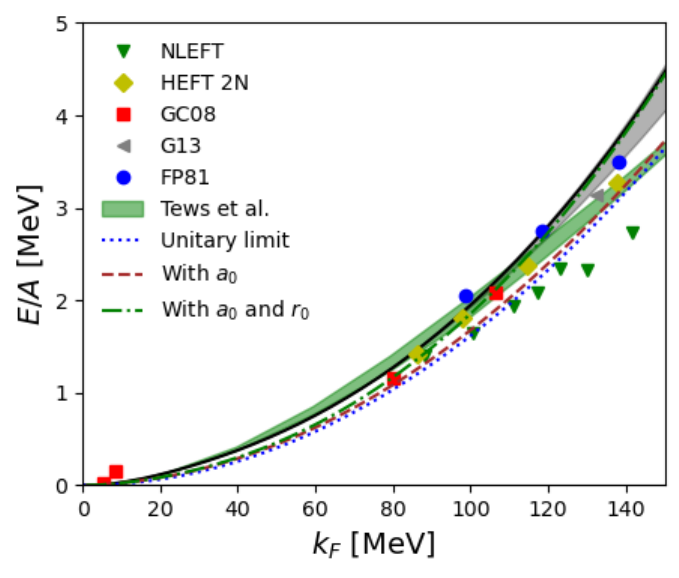}
\caption{{\small We show $E/A$ from \cite{Alarcon:2022vtn} (black solid line and grey band), and the results for NLEFT \cite{Epelbaum:2009rkz} (green downwards triangles), HEFT 2N \cite{Wlazlowski:2014jna} (yellowish green diamonds), the two quantum Monte Carlo
calculations of Gezerlis and Carlson \cite{Gezerlis:2007fs} (red squares) and Gezerlis {\it et al.} \cite{Gezerlis:2013ipa} (gray left-pointing triangles),  the variational one of
Ref.~\cite{Friedman:1981qw}} (blue circles), and N$^3$LO chiral EFT results \cite{Tews:2012fj} (green band). 
   \label{fig.240613.1}}
 \end{figure}

The resulting $E/A$ from \cite{Alarcon:2022vtn} is compared in Fig.~\ref{fig.240613.1}  with other non-perturbative calculations based on variational techniques \cite{Friedman:1981qw}, on several Quantum Monte Carlo approaches \cite{Wlazlowski:2014jna,Gezerlis:2007fs,Gezerlis:2013ipa}, on lattice EFT \cite{Epelbaum:2009rkz}, and from N$^3$LO chiral EFT \cite{Tews:2012fj}. The EoS of Ref.~\cite{Alarcon:2022vtn} is given by the black solid line with the gray band reflecting its estimated uncertainty. Except for the lattice EFT calculation \cite{Epelbaum:2009rkz}, we see a remarkable good agreement between these calculations. The only noticeable fact is a trend in \cite{Alarcon:2022vtn} to produce slightly larger values of $E/A$ for $n\gtrsim 0.01~$fm$^{-3}$ or $k_F\gtrsim 135$~MeV. This is also the case when compared with the perturbative chiral EFT calculations of Refs.~\cite{Tews:2012fj,Drischler:2020yad}, cf. Fig.~\ref{fig:LBandnolog} below.

The EoS (pressure $P$ vs energy density $\varepsilon$) of pure neutron matter is extracted from the knowledge of $E/A$ as a function of $n$, by using the relations
\begin{equation}\label{eq:densityenergy}
  \varepsilon= n\left(M_N+\frac{E}{A}\right)~,  
\end{equation}
and that for $T=0$, 
\begin{equation}\label{eq:firstlaw}
  P= n^2\,\frac{d(E/A) }{d n}  
\end{equation}
where $M_N$  is the nucleon mass.

Reciprocally, for each point of the EoS with $P(\varepsilon)$ assumed to be known, we obtain the corresponding values of $n$ and $E/A$ by using a discretized version of Eqs.~\eqref{eq:densityenergy} and \eqref{eq:firstlaw}, with the use of dense enough partitions: 
\begin{eqnarray}
n_{i+1} &=& \frac{\varepsilon_{i+1}}{M_N+(E/A)_{i+1}}\label{eq:discrenumberdensity}\\
P_i &=& n^2_i \frac{(E/A)_{i+1}-(E/A)_{i-1}}{n_{i+1}-n_{i-1}}
\label{eq:discrefirstlaw}.
\end{eqnarray}
These two equations are employed to solve for $(E/A)_{\rm{i+1}}$  and $n_{\rm{i+1}}$ given the earlier two points $i$ and $i-1$, and having at hand $P(\varepsilon)$. Namely, the resulting explicit expressions are:
\begin{align}
    \label{240614.1}
(E/A)_{i+1}&=\frac{\Delta_i-2M_N n_i^2+\sqrt{\Sigma_i}}{2n_i^2}~,\\
n_{i+1}&=\frac{-\Delta_i+\sqrt{\Sigma_i}}{2P_i}~,\nonumber\\
\Delta_i&=\left\{(E/A)_{i-1}+M_N\right\}n_i^2-n_{i-1}P_i~,\nonumber\\
\Sigma_i&=\Delta_i^2+4n_i^2 P_i\varepsilon_{i+1}~.\nonumber
\end{align}
 
 Furthermore, from the grand potential ensemble we calculate the baryon chemical potential $\mu$ for each point using the Euler equation
\begin{equation}
    \mu_i= \frac{\varepsilon_i+ P_i}{n_i} \ .
\label{eq:discrepotential}    
\end{equation}

\subsection{Interpolation process}
\label{sec.240614.1}

Starting from the obtained values for ($\varepsilon,P,E/A, n,\mu$) in the low-density region, we interpolate to higher densities using the interpolation procedure developed in \cite{LopeOter:2019pcq}. 
With this method, the first step is to establish the allowed region based on the conditions of causality and monotonic behavior. In terms of the sound velocity squared $c^2_s$, 
\begin{align}
    \label{240614.2}
c_s^2&=\frac{\partial P(\varepsilon)}{\partial \varepsilon}~,
\end{align}
these conditions imply that 
$0\leq c^2_s\leq 1$.  Along the work, we will also indicate when we take another maximum value for the sound velocity,   $c_s^2 \leq$ 0.781, according to Refs.~\cite{Hippert:2024hum,Tang:2024jvs}.

 Inside this region, we construct a grid of $1000 \times 1000$ candidate points ($\varepsilon, P$) representing potential EoS for the neutron matter. 
For each grid point $\varepsilon$, we select a value of $P$ of the grid by slowly increasing the slope at the beginning of the grid, and accelerating the slope growth with the energy density with varying values for each EoS, until the maximum slope $c_s^2 \leq 1$ is obtained, thus ensuring the principles of stability and causality. We also account for possible phase transitions (PTs) by increasing $\varepsilon$  while $P(\varepsilon)$ is held fixed. 
Moreover, we also calculate for each point of the EoS the value of $\mu$ by means Eqs.~\eqref{eq:discrenumberdensity}--\eqref{eq:discrepotential}, ensuring the thermodynamic consistency principle and the causality of speed of sound in the  $(n,\mu)$ plane. There, the stability and causality conditions read
\begin{align}
\label{240614.4}
c_s^{-2}&=\frac{\mu}{n}\frac{\partial n(\mu)}{\partial \mu}\geq 1~.
\end{align}
In addition, in each interpolation, we display the bands related to the integral constraints \cite{Komoltsev:2021jzg}, corresponding to the maximum and minimum $P(\varepsilon)$ values of the high matching band with pQCD. These bands are built according the procedure of Ref.~\cite{Komoltsev:2021jzg}. See this reference for the derivation and explicit expressions of the limiting curves $\{\varepsilon_{\text{max}}(\mu),P_{\text{min}}(\mu,n_{\text{max}}(\mu)\}$ and $\{\varepsilon_{\text{min}}(\mu),P_{\text{max}}(\mu,n_{\text{min}}(\mu)\}$, with $M_N\leq \mu<\mu_H$, and $\mu_H$ the upper limit for the chemical potential in the interpolation process. 

Following these constructions, we proceed with two interpolations: {\it i)} From the upper limit of applicability of the EoS \cite{Alarcon:2022vtn} in density, $n_L$, up to the saturation density $n_s$. {\it ii)} From $n_s$  up to the pQCD limit of around $40 n_s$, with a highest chemical potential $\mu_H=2600\pm 8$~MeV. In this way, for the first extrapolation we are able to restrict the information at $n_s$ from experimental information on nuclear reactions, and for the second one the constraints come by requiring the reproduction of the pQCD limiting values. In addition, we have also at our disposal the use of experimental data from astrophysics on neutron stars.

\bigskip 

\subsection{First interpolation,  $ n_s \geq n\geq n_L$}
\label{sec.firstinter}

   For the first interpolation, the experimental information  that we take in order  to constrain the EoS at $n_s$   stems  from  neutron skin measurements in ${}^{208}$Pb by  PREX-II  \cite{PREX:2021umo} and ${}^{48}$Ca by CREX  \cite{CREX:2022kgg} experiments, leading to values for the symmetry energy, $S_0$, and its slope, $L$, at the saturation density $n=n_s$. Here, the symmetry energy as a function of density is defined as $E_{\text{sym}}(n)=(E/A)_{\text{PNM}}(n)-(E/A)_{\text{SNM}}(n)$, and its slope $L_{\text{sym}}(n)=3n \partial E_{\text{sym}}(n)/\partial n$. In the following, we denote by $S_0=E_{\rm sym}(n_s)$ and $L=L_{\rm sym}(n_s)$.

From PREX-II \cite{PREX:2021umo} Ref.~\cite{Reed:2021nqk} deduces the values:
\begin{eqnarray}
  \label{240719.2}
     &  S_0= 38.1 \pm 4.7~{\text{MeV}} \\
     &  L = 106 \pm 37~{\text{MeV}}.
\end{eqnarray}
In turn,  Ref.~ \cite{Reed:2023cap} from the CREX experiment \cite{CREX:2022kgg} distinguishes between two scenarios and gives two possible values for $L_{\text{sym}}$:
\begin{align}
\label{240615.1}
\text{Scenario 1:} ~   &  ~ L = 110 \pm 40~\text{MeV}~,\\
\text{Scenario 2:} ~    & ~ L = 19 \pm 19~{\text{MeV}}~,\nonumber
\end{align}
and not providing any independent value for $S_0$.

Now, our point is to use this information to constraint the energy per particle and pressure for PNM at $n_s$, by also taking into account that at the equilibrium $P_\text{SNM}(n_s)=0$, and the phenomenological values 
$(E/A)_{\text{SNM}}(n_s) = -16.0 \pm 0.5$ MeV and $n_s = 0.16 \pm  0.01$~fm$^{-3}$ \cite{Bender:2003jk}. Therefore, $(E/A)_{\rm PNM}(n_s)$ and $P_{\rm PNM}(n_s)$ can be restricted within experimentally established intervals,  
\begin{align}
  \label{240719.1}
  (E/A)_{\rm PNM}(n_s) &= E_{\rm sym}(n_s) + (E/A)_{\rm SNM}(n_s)\\
  &\in (16.9,27.3)~\text{MeV} ~, \nonumber\\
     P_{\rm PNM}(n_s) &= \frac{1}{3} n_s L_{\rm sym}(n_s) ~.\nonumber
\end{align}
For the maximum value $  L_{\rm sym}(n_s) = 143$ MeV, we obtain a pressure at saturation density $P_{\text{PNM}}(n_s)=7.62$ MeV/fm$^{3}$. By the Euler relation Eq.~\eqref{eq:discrepotential} these values give us  the largest limit for a $\mu_H=1014.5~$MeV used in the construction of the allowed regions within the $(\varepsilon,p)$ plane following Ref.~\cite{Komoltsev:2021jzg}. 

Selecting a minimum value of $P_{\text{PNM}}(n_s)$ from CREX measurements is a much more complicated task, as clearly reflected in Eq.~\eqref{240615.1}.  Different Bayesian analyses \cite{Zhang:2022bni,Reed:2023cap,Koehn:2024set} found that the two experimental results are incompatible with each other at $68 \%$ confidence level, but compatible at $95\%$ confidence level. We first select the lowest central value of \cite{Reed:2023cap} $L_{\text{sym}}=19$~MeV, that implies $P_{\text{PNM}}(n_s)\approx 1.01$ MeV/fm$^{3}$.

Because of this limitation to clearly settle the low limit for $L$ from CREX, we have moved on 
to the Unitary Gas Conjecture (UGC) to constraint $E/A$ from below \cite{Tews:2016jhi,Lattimer:2023xjm}. In this model, the energy per particle of a unitary Fermi gas of neutrons $(E/A)_{UG}$ is taken  as a lower bound \cite{Tews:2016jhi,Lattimer:2023xjm} for $E/A$ in PNM, and it is supported by many explicit calculations. In a unitary Fermi gas the $S$-wave scattering length is infinity and no other parameters in the effective range expansion nor higher partial waves are included in the neutron-neutron interactions. The resulting expression for $(E/A)_{UG}$ is universal and given by $(E/A)_{UG}=\xi (E/A)_F$, where $\xi$ is the
Bertsch parameter, $\xi=0.370(5)(8)$ \cite{Zuern:2013}, 
and $(E/A)_F$ is $E/A$ for a free Fermi gas. By applying Eq.~\eqref{eq:firstlaw} we then have the values $(E/A)_{\rm UG}(n_s)=12.7\pm 0.3~$MeV, $\varepsilon_{\rm UG}(n_s)= 152.46\pm 0.05$~MeV/fm$^3$, $P_{\rm UG}(n_s)=1.33\pm 0.04$~MeV/fm$^{3}$ and $L_{\rm UG}=24.9\pm 0.7$~MeV. By applying Eq.~\eqref{eq:discrepotential} these values imply a  chemical potential at $n_s$ equal to $\mu_H=960.6\pm 0.6$~MeV. This  chemical potential is the smallest limit for a  $\mu_H$ to be used in the construction of allowed regions of EoS in the $(\varepsilon,p)$ plane from Ref.~\cite{Komoltsev:2021jzg}. 
\bigskip

\subsection{Second interpolation for higher densities, $n>n_s$}
\label{sec.secondinter}

For the second extrapolation up to  very high densities, the EoS can also be computed directly from the fundamental theory of Quantum Chromodynamics (QCD). This theory  becomes perturbative at asymptotically high energies, which are reached either at high temperatures or at high densities, at about 20--40~$n_s$ in baryon number density, well above the densities reached in NS \cite{Kurkela:2009gj,Kurkela:2016was,Gorda:2021kme,Gorda:2023mkk}.  
However, the existence of 
this pQCD  regime implies a limit to the pressure, energy density, and baryon chemical potential at the pQCD matching point. This fact is certainly very interesting because it provides constraints on the EoS and creates a  bounded region through which valid EoS $P(\varepsilon)$ must pass. The partial N$^3$LO pQCD results of \cite{Gorda:2021kme} shows that this regime is valid at  $\mu_{pQCD}=2.6$~GeV and these results still depend on a renormalization scale $X$  \cite{Fraga:2013qra,Komoltsev:2021jzg}.\footnote{$X=3\bar{\Lambda}/\mu$,  with $\bar{\Lambda}$ the $\overline{\text{MS}}$ renormalization scale} 
All the points of the EoS can be connected to the pQCD values, ($\mu_{pQCD}, \varepsilon_{pQCD}, P_{pQCD}$), at any point inside the band corresponding to $\mu_{pQCD}=$ 2.6 GeV and the renormalization scale $X \in [1,4]$ \cite{Komoltsev:2021jzg, Gorda:2021kme}. This connection has to be made by some causal, stable and thermodynamically consistent interpolation, taking into account the squared speed of sound in the plane ($P,\varepsilon$), $c_s^2= \partial P/\partial \varepsilon \leq 1$, as well as in the plane $(n,\mu)$, $c_s^2=(n/\mu)(\partial  \mu/\partial n \leq 1)$, cf. Eqs.~\eqref{240614.2} and \eqref{240614.4}, respectively.

There are different ways to take into account the results of pQCD at high chemical potential, $\mu_H\approx 2.6~$GeV. Some models \cite{Somasundaram:2022ztm} terminate the extension of the EoS  at a certain $n_{\text {term}}=n_{c}$ (where $n_{c}$ is the central density number of the NS with the maximum mass possible for a given EoS, when the TOV equations are solved for a non-rotating NS), but others models extrapolate their EoS up to a much higher density $n_{\text term}\approx 10 n_s$ \cite{Gorda:2022jvk,Brandes:2023hma}. The latter references apply the limits for $P(n)$ at every $n$ derived from the integral constraints \cite{Komoltsev:2021jzg}:

\begin{equation}
\int_{\mu(n)}^{\mu_{\rm H}} n(\mu) d\mu =P_{\rm H}-P(n)~ ,
\end{equation}
where  the subscript $H$ refers to  matching with the pQCD results. Then $P(n)$ should be within the ranges $P_{\text{min}}$ and $P_{\text{max}}$ already discussed.

In our interpolation, for the process of matching with pQCD it is important to take into account  the value of the so-called critical chemical potential, $\mu_c$, that corresponds  to the pressure at the intersection point in the $(n,\mu)$ plane between the causal lines and the lower boundary of the branch connecting with pQCD at $\mu_H$. Its expression is given by  \cite{Komoltsev:2021jzg}: 
\begin{equation}
\label{eq:muc}
\mu_c = \sqrt{\frac{\mu_L \mu_H ( \mu_H n_H-  \mu_L n_L-2 \Delta p )}{\mu_L n_H - \mu_H n_L} }.
\end{equation}
where the subscript  
 $L$ denotes the low-density
limit for this second extrapolation starting at $n_s$. 
If one considers velocities of sound $c_s^2\leq$ 1, this upper value cannot be held for chemical potentials $\mu\geq \mu_c$ in order to enter pQCD with  $\mu_{H}=$2600~MeV. Thus, we maintain the maximum velocity until we reach a potential $\mu_0 < \mu_c$ and, at this point, we link to pQCD by performing a long PT to the energy density value that allows reaching the pQCD band with a slope $c_s^2 \leq$ 1/3. Other possibilities are also explored as, for example, instead of such a long PT we also allow a shorter one followed by an increase in $c_s^2\leq 1$,  and then  repeating this process a few times, the number depending on the EoS within the limits settled by the allowed regions constructed from the first principles. After the final PT  the pQCD limit is approached in the way just explained. 
In this work,  we typically have that $\mu_0 \approx 2200~\text{MeV}<\mu_c=2250~$MeV. For a lower slope case, such as  $c_s^2 \leq$ 0.781, $\mu_0 \approx$ 2020~MeV, around a 10$\%$ lower than when allowing $c_s^2\leq 1$. In both cases,  $n(\mu_0)$ is less than or equal to the central density point in the star, when the TOV equations are solved.

Apart from these theoretical constraints coming from pQCD at very high densities, the intermediate density region (up to several times $n_s$, more details are given below) is further constrained with data from measurements on the mass-radius and tidal deformability of NSs. 

Regarding masses, the heaviest neutron stars derives mainly from
precise Shapiro time delay measurements of pulsars orbiting in binary systems, such as PSR J1614–2230  \cite{Demorest:2010bx,Fonseca:2021wxt}, PSR J0348+0432 \cite{Antoniadis:2013pzd} and PSR J0740+6620 \cite{Cromartie:2019kug,Fonseca:2021wxt}, with reported masses $M \gtrsim 2M_\odot$. Furthermore, the heaviest neutron star observed is the black-widow pulsar PSR J0952-0607,which was recently reported by \cite{Romani:2022jhd} with a mass at 68\% confidence level.

PSR J0952-0607
\begin{equation}
     M= 2.35 \pm 0.17 \,M_\odot
 \label{eq:J0952}    
\end{equation}

There are others sources of massive NS, such as the hypothesis that the remnant in GW170817, with a total mass $M=2.82^{+0.47}_{-0.09}\,M_\odot$, collapsed to a black hole by data from the short gamma-ray burst GRB170817A \cite{LIGOScientific:2017ync}. 
Another  example of massive NS is GW190425 \cite{LIGOScientific:2020aai}, a GW signal  candidate for black hole-neutron star merger, with a probable mass of the neutron star $M=2.31^{+0.56}_{-0.26}\,$$M_\odot$.

On the other hand, the observation of a small  NS in the supernova remnant HESS J1731-347 has been reported \cite{2022NatAs...6.1444D}, with  mass and radius at the 68$\%$ confidence level:
\begin{equation}
     M=0.77^{+0.20}_{-0.17} \,M_\odot,\hspace{1cm}  R=10.4^{+0.86}_{-0.78}~\text{km}
\label{eq:HESS}     
\end{equation}

Additionally, masses and  radii of NSs can be obtained from X-ray profiles of rotating hot-spot patterns measured with the NICER telescope. In this way, there are results for three NSs at 68$\%$: 

\begin{itemize}
    \item PSR J0030+0451:  

    \begin{align}
    \label{240616.1}
M&=(1.34^{+0.50}_{-0.16})\,M_\odot~, ~R=(12.71^{+1.14}_{-1.19})\,\text{km~ \cite{Riley:2019yda}}
    \\
  M&=(1.4\pm0.05)\,M_\odot~,~ R=(13.02^{+1.24}_{-1.06})\,\text{km~ \cite{Miller:2019cac}}~.\nonumber
\end{align}

    \item PSR J0740+6620: 
    
\begin{align}  
\label{240616.2}
M&=(2.08 \pm 0.07)\,M_\odot~,~ R=(12.92^{+2.09}_{-1.13})\, \text{km ~\cite{Dittmann:2024mbo}}
    \\
M&=(2.073\pm 0.069)\,M_\odot~,~ R=(12.49^{+1.28}_{-0.88})\, \text{km~  \cite{Salmi:2024aum}}
  ~.\nonumber
    \end{align}

    \item PSR J0437-4715: 

\begin{align}
\label{240616.3a}
M&=(1.418 \pm 0.037)\,M_\odot~, ~R=(11.36^{+0.95}_{-0.63})\,\text{km~\cite{Choudhury:2024xbk}} .
\end{align}
\end{itemize}

Another astrophysical input is  the tidal deformability of NS in binary mergers \cite{TheLIGOScientific:2017qsa,Abbott:2018exr,LIGOScientific:2018hze}. An upper bound on the dimensionless binary tidal deformability parameter $\bar{\Lambda}\leq$ 720 (low-spin priors) has been obtained from GW170817 \cite{LIGOScientific:2018hze}.  A $1.4\,M_\odot$ NS tidal deformability was deduced in \cite{Abbott:2018exr} with the result:
\begin{equation}
     \Lambda_{1.4}= 190^{+390}_{-120}
\end{equation}

Further investigation on the GW170817 event together with electromagnetic signals \cite{Fasano:2019zwm} reported  the following data  for the individual NSs in the binary:

\begin{align}
\label{240614.6}
M&=1.46^{+0.13}_{-0.09}\, M_\odot\,, ~ \Lambda=255^{+416}_{-171}~,\\
 M&=1.26^{+0.09}_{-0.12}\, M_\odot\,,~  \Lambda=661^{+858}_{-375}~.\nonumber
\end{align}

The EoS is also constrained by forbidding a PT  below $2.5 n_s$.
This constraint comes from the in-medium corrections of the quark condensate performed in \cite{Lacour:2010ci,Meissner:2001gz}. 
This quantity plays the role of an order parameter in nuclear matter, and depends on the value of the pion-nucleon sigma term ($\sigma_{\pi N}$). 
Higher values of $\sigma_{\pi N}$ favor the vanishing of the in-medium quark condensate at lower densities (as a first approximation the in-medium quark condensate  depends linearly in $\sigma_{\pi N}$). 
For the largest phenomenological values $\sigma_{\pi N} \approx 60$~MeV \cite{Alarcon:2011zs, Hoferichter:2015dsa}, 
one could have a vanishing of the in-medium quark condensate  at $\approx 2.5 n_s$. We just mention by passing that there is an ongoing tension sustained in time \cite{Alarcon:2021dlz} between these phenomenological determinations and the ones from lattice QCD, which generate smaller values of $\sigma_{\pi N}$ somewhat above 40~MeV, like the recent calculation $\sigma_{\pi N}=43.7\pm 3.6$~MeV \cite{Agadjanov:2023efe}.

\section{Results}
\subsection{Interpolation up to the saturation density}
\label{sec.240616.1}

\begin{figure}
     \centering
     \includegraphics[width=0.49\textwidth]{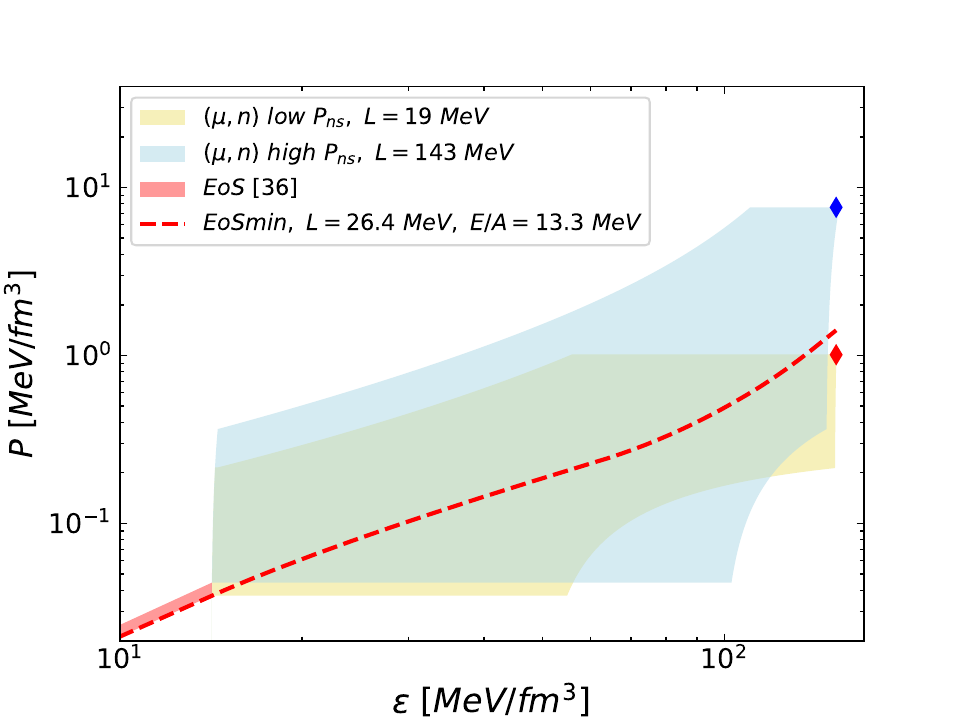}
\caption{{\small Bounding areas of validity for PNM EoS by extrapolating the EoS of Ref.~\cite{Alarcon:2022vtn} (reddish band) from $n_L$ up to $n_s$. The blue and red diamonds correspond to the highest chemical potentials used in the construction of Ref.~\cite{Komoltsev:2021jzg}, as explained in Sec.~\ref{sec.240614.1}, for $L=143~$MeV and 19~MeV, respectively. They give rise to the palish blue and yellow areas, in this order. The value $L\approx 25~$MeV is fixed by the UGC (as given above), closely approximated by the red dashed line.}
\label{fig:Initialband}}
 \end{figure}

The EoS from Ref.~\cite{Alarcon:2022vtn} is plotted in the left corner of Fig.~\ref{fig:Initialband} by the light green band up to its limit $n_L=0.1 n_s$. Then, we proceed along the lines of the  extrapolation method detailed in Sec.~\ref{sec.240614.1} for the low and intermediate density region up to $n=n_s$.

The EoS corresponding to the upper and lower limit points of the integral constraints corresponding to the first interpolation, taking as the highest value for the pressure at $n_s$ either $P_s=$1.01~MeV/fm$^{3}$ ($L_{\rm sym}=19$~MeV) or  7.62~MeV/fm$^{3}$  ($L_{\rm sym}=143$~MeV), are shown in Figure \ref{fig:Initialband} by the palish yellow and blue areas, respectively. Growing  the EoS through the grid points of the first   interpolation, we find that the minimum EoS   (red dashed line)  compatible with $(E/A)_{UG}$ presents $P_s=1.4~$MeV/fm$^{3}$ at $n_s$, giving $L=26.4$ MeV and $E/A=13.3$ MeV. This value, in accordance with UGC, is smaller than the interval $(E/A)_{\rm PNM}(n_s)\in(16.9, 27.3)$~MeV  given in  Eq.~\eqref{240719.1} from the phenomenological values of $S_0$, $L$, $(E/A)_{\rm SNM}(n_s)$ and $n_s$ there discussed. 
  
\begin{figure}
     \centering
     \includegraphics[width=0.49\textwidth]{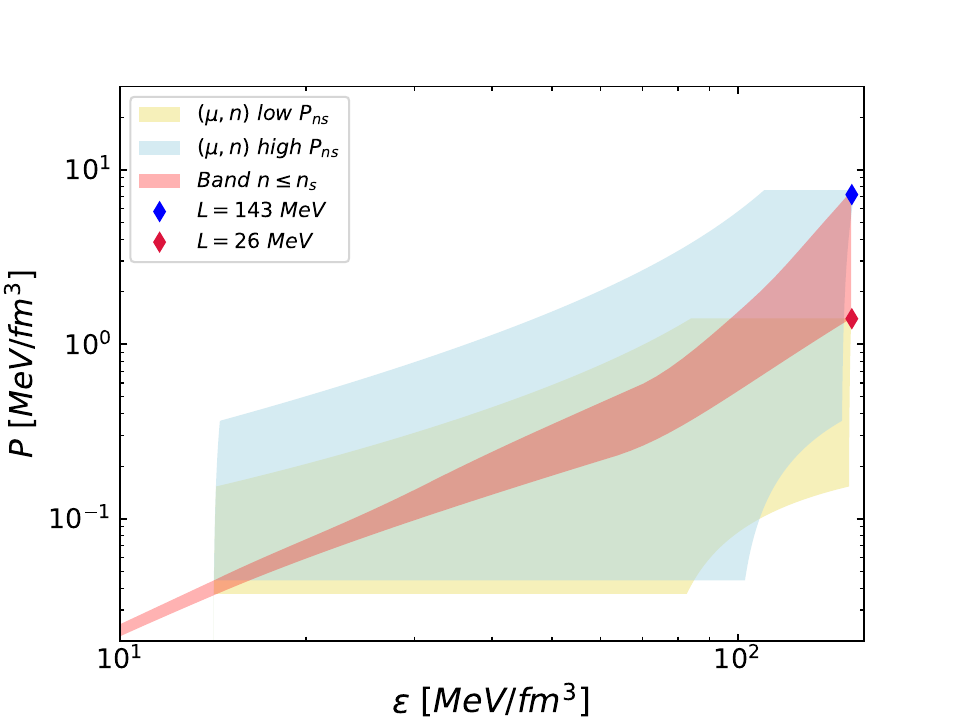}
 \caption{{\small Extrapolation from the upper limit of validity of the EoS \cite{Alarcon:2022vtn} up to $n_s$, $n_L<n<n_s$.  The same bounding bands of acceptable areas as in Fig.~\ref{fig:Initialband} are shown but now the read diamond is fixed from the red dashed line in Fig.~\ref{fig:Initialband}. The resulting  band of EoS's  from the extrapolation up to $n_s$ is the narrower red area.  } 
     \label{fig:LBand}}
 \end{figure}

Finally, we take  $P(n_s)$  from the red dashed line in Fig.~\ref{fig:Initialband} as the lower limit at $n_s$, and do the interpolation matching at $n_s$ within the extreme values  $P_{\rm high}=7.62$~MeV/fm$^{3}$ and $P_{\rm low}=1.41$~MeV/fm$^{3}$.  The chemical potentials $\mu_H$ corresponding to these points are $\mu_H=1014.5$ and $961.7$~MeV, respectively. The allowed regions of EoS's stemming from the blue and red diamonds, corresponding to the former and latter pressures, respectively, are shown in Fig.~\ref{fig:LBand} by the palish blue and yellow areas, in this order.   For the interpolation process,  we construct a 100-point EoS using a 1000 $\times$ 1000-point grid to achieve better control over the slope. We construct the EoS's from the low-density band to the higher-density one by raising smoothly $c_s^2$,  keeping 
$dc_s^2/d\varepsilon >0$, with this derivative being slightly larger or smaller depending on the point in the grid, until $n_s$ is reached, where the interval of pressures between the extreme ones is filled. 
This gently rise in $c_s^2$  also avoids having to introduce PTs at too low densities because of touching the borders of the allowed region. 
We display  the final band results for $P(\varepsilon)$ up to about $n_s$ in Fig.~\ref{fig:LBand}. The value of $L$ for the EoS that runs along the upper and lower limits of the total band corresponding to the red area in the figure are also given in the keys of the figure.

\begin{figure}
     \centering
     \includegraphics[width=0.49\textwidth]{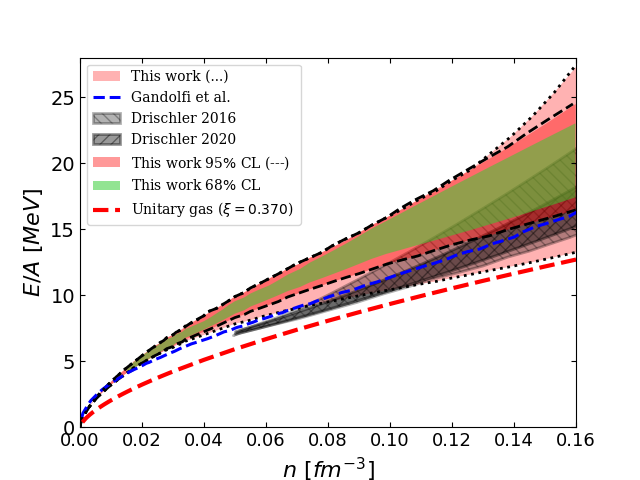}
 \caption{{\small We show our final band of  EoS's up to $n_s$ in three cases: The pink area corresponds to the direct extrapolation up to $n_s$, with its borders separated by dots. Once astrophysical constraints are taken into account the allowed area shrinks and it is depicted in green (68\% CL) and in orange (95\% CL), with the borders for the latter signaled by dashed lines. We also compare with other results:  The blue dotted-dashed line is the Monte Carlo calculation of Ref.~\cite{Gandolfi:2011xu}. The many-body perturbation calculations from chiral potentials are given by the light  \cite{Drischler:2016djf} and darker \cite{Drischler:2020hwi}  gray areas. The dashed one is the unitary limit for a Fermi gas. }
     \label{fig:LBandnolog} }
 \end{figure}

Our band of EoS's at this stage is also shown by the pink area in Fig.~\ref{fig:LBandnolog}, compared with other calculations from quantum Monte Carlo \cite{Gandolfi:2011xu} (blue dotted-dashed line), and the many-body perturbative methods with chiral potentials of Refs.~\cite{Drischler:2016djf,Drischler:2020hwi} (gray areas). In addition we also show by the red dashed line the unitary gas limit result. 

\subsection{Interpolation up to the pQCD regime}
\begin{figure}
     \centering
  \includegraphics[width=0.49\textwidth]{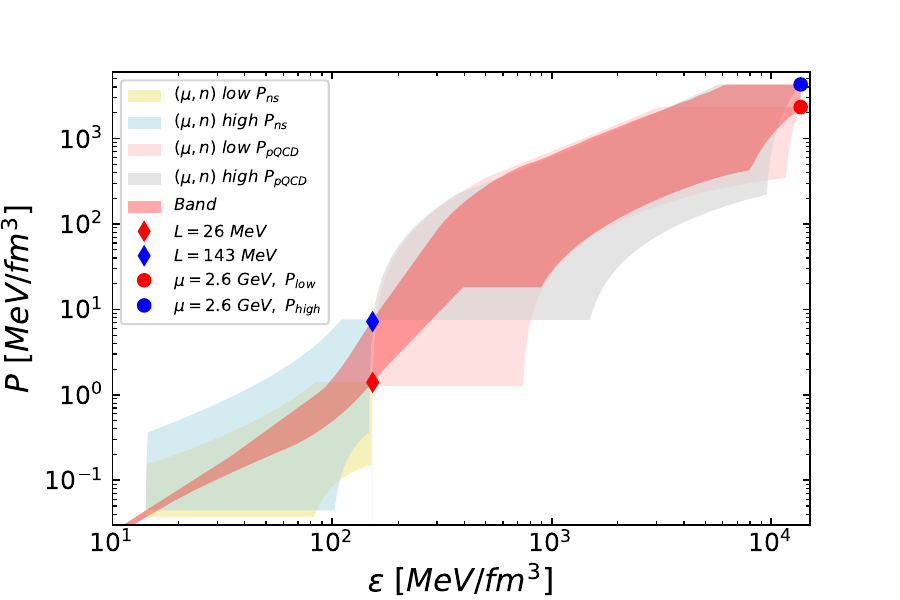}
 \caption{{\small Band  obtained by interpolation from low to high densities, from the EoS \cite{Alarcon:2022vtn} up to pQCD. This extrapolating band is limited only by first principles (causality and thermodynamical consistency), pQCD and from measurements of nuclear parameters. }
     \label{fig:bandnofilter} }
 \end{figure}

\subsubsection{Without astrophysical observables}
We discuss here the practical realization of the method explained in Secs.~\ref{sec.240614.1} and \ref{sec.secondinter}.   
We first apply the construction of Ref.~\cite{Komoltsev:2021jzg} and determine the areas in the $(\varepsilon,P)$ plane inside which a viable EoS can interpolate between $n_s$ and the pQCD region at around $40 n_s$. At the lowest density we have the band determined in Sec.~\ref{sec.240616.1} at $n_s$, within the two extreme pressure values 1.4 and 7.6~MeV/fm$^{3}$, and shown in Fig.~\ref{fig:LBand}. At the highest density there is the band from the N$^3$LO pQCD EoS \cite{Gorda:2021znl,Gorda:2021kme} at $n \approx 40n_s$, within the extreme pressures  2334 and 4384~MeV/fm$^{3}$. 
In the process these areas for viable EoS's  connect the lowest and highest pressures at the extreme densities taken for the interpolation. These areas, and those already determined in Sec.~\ref{sec.240616.1} for the extrapolation between $n_L$ and $n_s$, are shown in Fig.~\ref{fig:bandnofilter}. A $10^3\times 10^3$ grid inside the region fulfilling Eq.~\eqref{240614.2} is built that in turn comprises these areas between $n_s$ and $40 n_s$ is built.   We then construct each point of the EoS ($n,E/A,\varepsilon,P,\mu$) by applying the minimum slope growth, so that it maintains  positive its derivative $dc_s^2/d\varepsilon>$0,   
with slightly larger or smaller values depending on the point and the EoS,  between  $n_s \lesssim n \lesssim 2.5n_s$.  In the interval $2 n_s\leq n\leq  2.5 n_s$ this derivative steadily increases for the stiffest EoS's that drive to preferred NS mass-radius relation according to present data, as we analyze in more detail  below.

From $n>2.5n_s$, in order to consider the possibility of PTs, we apply the condition $dc_s^2/d\varepsilon \geq$0, allowing that it can be equal to zero. This derivative is smoothly increased  until $c_s^2$ reaches its limit value, which is then kept until a  chemical potential $\mu_0<\mu_c=2250~\text{MeV}$, for the causal value $c_s^2\leq 1$, and $\mu_0\lesssim 2040~\text{MeV}<\mu_c$ for the upper bound $c_s^2\leq 0.781$ \cite{Hippert:2024hum,Tang:2024jvs}. For the case with PTs for $\mu<\mu_0$  we apply the limit value for $c_s^2$ between PTs in order to have a   sufficiently stiff EoS. As discussed in Sec.~\ref{sec.secondinter}, 
at the energy density  corresponding to this high chemical potential,  varying for each EoS, we consider different possibilities. One of them consists of making  a long PT with an energy-density jump such that, from this PT onwards, pQCD is reached with $c_s^2\to 1/3$ starting with $c_s^2\approx 0.3$. Other interpolations involve a shorter PT followed by an increase in $c_s^2\leq 1$, and then  repeating this process again for a few times, depending of each EoS.  After the final PT the pQCD limit is reached as explained before.

 The resulting EoS band  for a NS that  follows from this interpolation procedure is depicted by the red area in Fig.~\ref{fig:bandnofilter}  from the EoS \cite{Alarcon:2022vtn}  to the core and extended beyond up to pQCD, constrained only by causality, thermodynamical stability, pQCD and  measurements from nuclear parameters. It is worth stressing that this band is not constrained by astrophysical observables and, as a result, it is convenient for testing GR and modify theories \cite{Lope-Oter:2023urz,Lope-Oter:2024egz}.  We notice that for the soft EoS's, $n=2.5 n_s$ is reached for low values of $P$ ($P<50$ MeV/fm$^3$), such that 
the EoS needs a specially fast  increase of $c_s^2$ after the first PT. They also  make other PTs for entering pQCD with $c_s^2\leq 1/3$.

\subsubsection{With astrophysical observables}

Next, we take into account astrophysical observables in order to constrain further the EoS and apply GR to calculate them. We use the values of Mass and Radius of pulsars: PSR J0952-0607, Eq.~\eqref{eq:J0952}, HESS J1731-347, Eq.~\eqref{eq:HESS}, PSR J0030+0451, Eq.~\eqref{240616.1}, PSR J0740+6620, Eq.~\eqref{240616.2}, and PSR J0437-4715, Eq.~\eqref{240616.3a}. 
Solving together the Tolman-Oppenheimer-Volkoff (TOV) equations for non rotating-NS and tidal deformability equations for a variety of central pressures, the mass-radius relation, and the tidal-deformability/mass-radius relations are obtained for each given EoS.

We  apply  the method of least squares by using these six independent mass measurements $M_i$ at known radius 
$R_i$ (except for J0952-0607, see below) to obtain the band of EoS's that minimizes the following $\chi^2$:
\begin{align}
\label{240616.3}
    \chi^2=\sum_{i=1}^6 \frac{(M_i-m(R_i))^2}{\sigma_i^2}
\end{align}
where the $i_{\text{th}}$ mass measurement $M_i$ is assumed to be Gaussian distributed with  known variance $\sigma_i^2$, while the theoretical calculation is represented by $m(R_i)$. 
All the $\sigma_i$ values used correspond to the 68$\%$ confidence interval values derived from each analysis. If for an event they are asymmetrical, we take its left/right 
value if the point $y_i$ is found to lie to the right/left of the theoretical band of values (a building up consistency criterion). For the case of  PSR J0952-0607 its radius is not provided, cf. Eq.~\eqref{eq:J0952}. Because of this, and regarding this pulsar, we can only demand that the EoS under consideration can generate NS masses large enough to accommodate its mass. Then, if the EoS considered generates a mass smaller than the upper bound for the mass of the PSR J0952-0607 ($M \geq2.53$ $M_\odot$) we apply the standard definition for the contribution of this event to the $\chi^2$.  If the highest allowed mass from this EoS is larger than the upper bound for the mass of the PSR J0952-0607  we do not add any contribution to the $\chi^2$ from this mass measurement. Then, there is window for more massive NS, such as in the event GW190425.

For each EoS, we calculate in the way explained the $\chi^2$  for the six measurements and  select those EoS's whose $\chi^2$ distribution reports a $p$-value $\geq$ 0.32 for  68$\%$ confidence level, and a $p$-value $\geq$ 0.05 for 95$\%$ confidence level. The set of EoS's with the former $p$-value makes the 68$\%$ band, while the set with the last $p$-value makes the 95$\%$ band. 

\begin{figure}[H]
     \centering
      \includegraphics[width=0.49\textwidth]{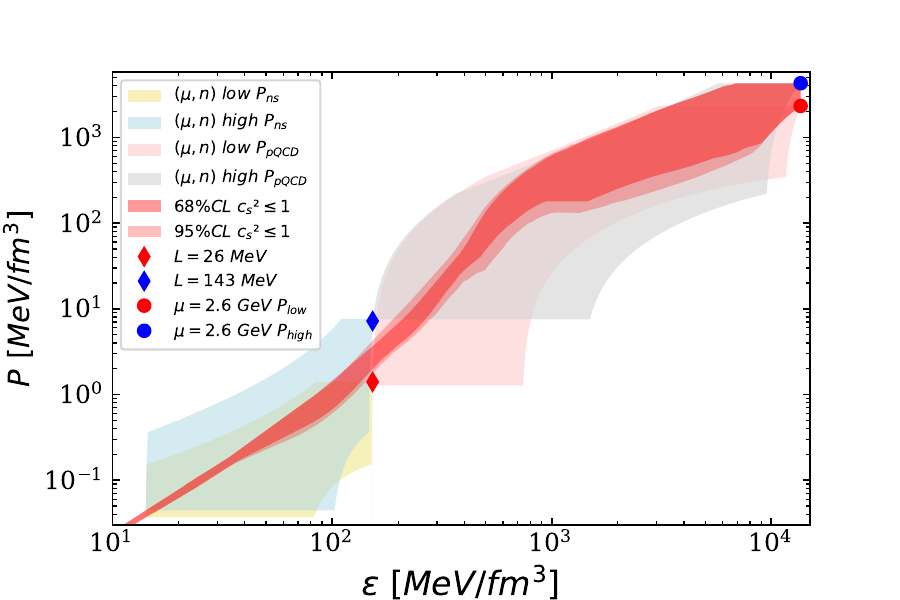}
     \includegraphics[width=0.49\textwidth]{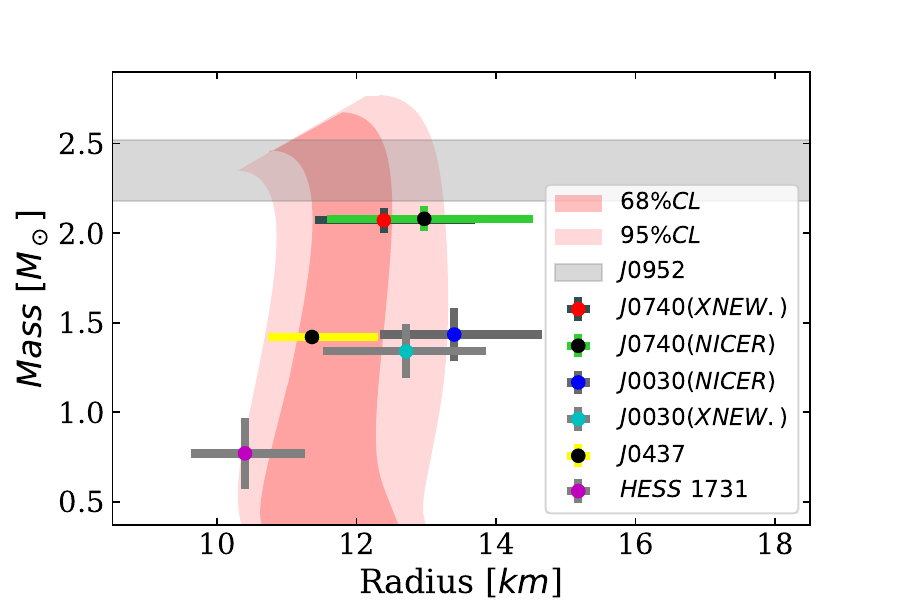}     
\includegraphics[width=0.47\textwidth]{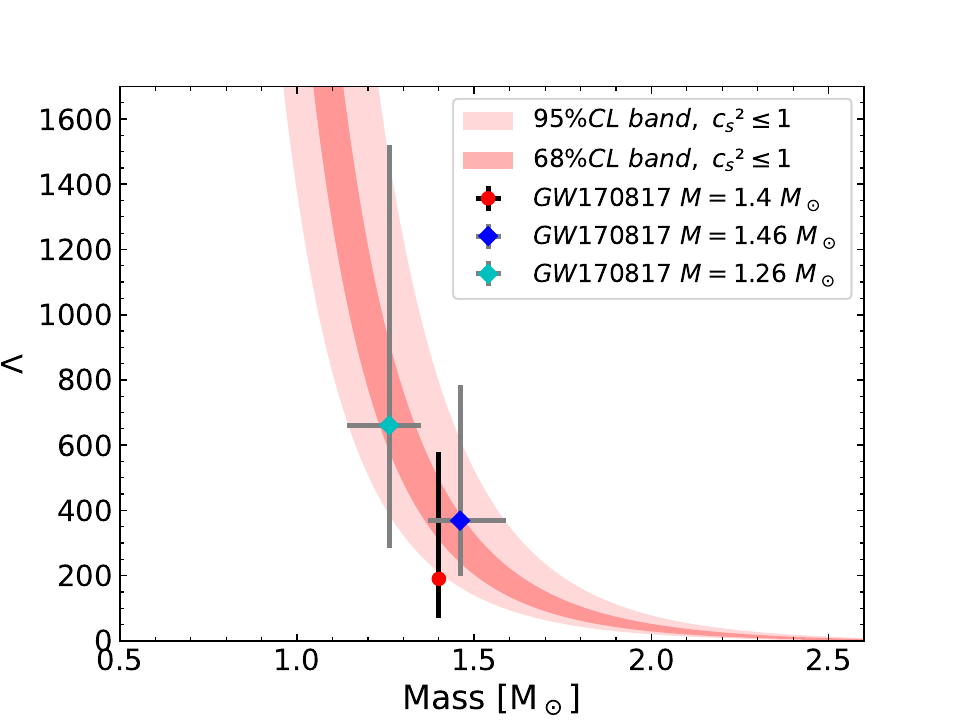}

\caption{{\small {\bf Top:} ($\varepsilon, p$) bands constrained  from astrophysical observables at 68$\%$ CL (red)  and 95$\%$ CL (pink) for $0\leq c_s^2 \leq 1$. {\bf Middle:}~ Mass-Radius diagrams for the EoS bands at the top. {\bf Bottom:}~ Tidal deformability-Mass relation for the EoS bands at the top. }
     \label{fig:completeband}}
 \end{figure}

The EoS bands  obtained in the case of maximum value $c_s^2 \leq 1$ are shown  at the top panel of Fig. \ref{fig:completeband} at the 68\% CL (red band) and 95\% CL (pink band).  The resulting bands are much narrower than those in Fig.~\ref{fig:bandnofilter} before taking into account the astrophysical observables. It is worth stressing that, in this way, we can also constrain the nuclear observables, providing sharper values for  the energy symmetry and its slope,

\begin{align}
\label{240616.5}
&    32.9\leq S_0 \leq 39.5~\text{MeV}\,; ~  
    37.3 \leq L\leq 69.0~\text{MeV}~  (68\% \,\text{CL})~,\\
 &   32.1\leq S_0 \leq 40.6~\text{MeV}\,; ~34.6 \leq L\leq 80.0~\text{MeV}\  (95\% \ \text{CL})\,.\nonumber    
\end{align}

The interval for the obtained symmetry energy $S_0$ at 68~$\%$ remarkably lies inside 1$\sigma$ PREX-II value, Eq.~\eqref{240719.2}, and its slope $L$ is almost at the one $\sigma$ level for both scenarios of CREX, Eq.~\eqref{240615.1}. Our range  for $S_0$ at the 68\% CL is  compatible  with $30-35~\text{MeV}$  \cite{Roca-Maza:2015eza}, $33^{+2.0}_{-1.8}$ \cite{Essick:2021kjb},   and $32\pm 1.7$~MeV  \cite{Lattimer:2023xjm}.

The calculated bands for the NS mass $M$ in units of $M_\odot$  as a function of its radius $R$ in km at the 98\% CL (red area) and 95\% CL (pink area) are shown in the middle panel of Fig.~\ref{fig:completeband}. The different experimental values considered and discussed in Sec.~\ref{sec.secondinter} are also indicated and explained in the keys of the figure. We see that the band within the one $\sigma$ region is narrow, rather vertical, and  centered for a radius value somewhat below 12~km.

In the bottom panel of Fig.~\ref{fig:completeband}  we show within the same pattern the obtained tidal deformability $\Lambda$ as a function of the NS mass in units of $M_\odot$ confronted the experimental determinations from the GW170817 event \cite{LIGOScientific:2017ync} for three masses, 1.26, 1.4 and 1.36~$M_\odot$. We see that our resulting band of values agrees very well with the central values determined for 1.26 and 1.46~$M_\odot$, and it is compatible within one $\sigma$ with the determination for $1.4~M_\odot$.

Regarding maximal central values for quantities of interest in a NS, the chemical potential reaches values of $ 1420\leq \mu_{\rm c,max}\leq 2220$~MeV at 68$\%$~CL, and within $ 1400 \leq \mu_{\rm c,max} \leq 2240$~MeV at 95$\%$~CL. With respect to the central baryon number density  $n_{c}$, it varies in the interval $5.0 n_s \leq n_{c} \leq 7 n_s$  at 68$\%$~CL,  and within $4.5 n_s \leq n_{c} \leq 7.4 n_s$ at 95$\%$~CL. For the energy per particle we find $ 280\lesssim (E/A)_{\rm c,max} \lesssim 400$MeV  and $ 280\lesssim (E/A)_{\rm c,max} \lesssim 430 $MeV at the 68$\%$ and 95\%~CL, respectively.  All PTs that we find occur for $M\gtrsim 2.1\,M_\odot$ both at 68 and 95 $\%$ CL, within the range of starting densities between $2.5 n_s \leq n \leq n_{c}$. Both  PTs with short and long phase-coexistence regions are found.  For $c_s^2 \leq 1$, short PTs appear for $2.5 n_s< n< 3.2n_s$ with $ 0.12< \Delta n/n<0.50$ (68$\%$ CL), and  for $2.5 n_s< n< 3.2n_s$ with $ 0.12< \Delta n/n<0.60$ (95$\%$ CL), and they typically require afterwards a steep rise in $c_s^2$ to confront well with NS mass-radius data. Because of this they could also stand for a second PT that starts at  densities less than or equal to the central density of the NS. The longest PTs found have a phase-coexistence region that extends over a range of densities $3.4 n_s\leq \Delta n \leq 19 n_s$ at 68$\%$ CL, and $3.2 n_s\leq \Delta n \leq 21 n_s$ at 95$\%$ CL, such that  $1 \lesssim \Delta n/n\leq 3.8$ at 68$\%$ CL and $1 \lesssim \Delta n/n\leq 4.5$ at 95$\%$ CL. Let us notice that  Refs.~\cite{Brandes:2023hma,Brandes:2023bob} conclude against the possible existence of PTs inside a NS with $M<2.1\,M_\odot$, in agreement with our results at 68$\%$ CL.

\begin{figure}[H]
     \centering
      \includegraphics[width=0.49\textwidth]{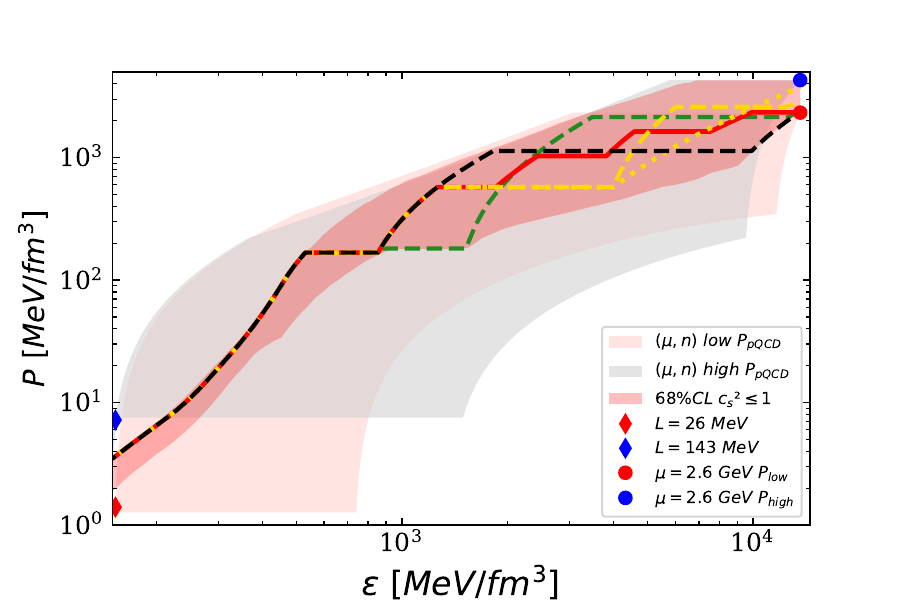}
     \includegraphics[width=0.49\textwidth]{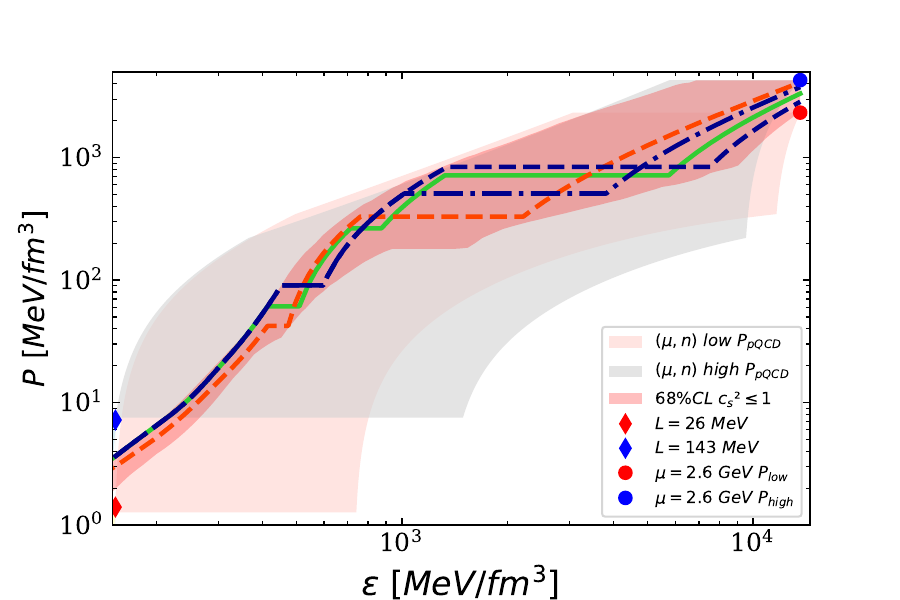}     

\caption{{\small All the EoS's considered in this plot are constrained from astrophysical observables at 68$\%$ CL for $0\leq c_s^2 \leq 1$. {\bf Top:} We plot 4 EoS's with the same first PT at $P=167~$MeV/fm$^3$ but reaching pQCD in distinct ways involving different number of PTs. 
 {\bf Bottom:}~  We show 3 EoS's with several PTs all of them starting at densities inside the NS.  For the two panels see the text for the meaning of the lines and more details.}}
     \label{fig:PTs}
 \end{figure}

The top panel in Fig.~\ref{fig:PTs} shows 4 EoS's at 68$\%$ CL with the same first PT at $P=167~$MeV/fm$^3$ and showing different ways to reach pQCD. For convenience, let us denote by $\tilde{\mu}$ the values of the chemical potential along the upper border line of the of the 68\%~CL region in the figure. Then, we consider EoS's with 2 PTs in total, such that a second PT at about $\tilde{\mu}$ takes place for the EoS represented by the dashed black line, while for the dotted gold line the second PT occurs below $\tilde{\mu}$. For 3 PTs in total we have the green line, with the second(third) PT happening below(about)  $\tilde{\mu}$, while for the gold dashed line the 2 extra PTs happen both below $\tilde{\mu}$. Finally, we also plot by the red line an EoS with 4 PTs in total, with all the extra PTs bellow $\tilde{\mu}$. All these EoS's enter pQCD with $c_s^2\leq$ 1/3. For the bottom panel of Fig.~\ref{fig:PTs} we depict 4 EoS's with several PTs all of them starting at densities inside the NS. The red and the two blue lines have 2 PTs, while the green line has 3 PTs. In between the blue dashed and dot-dashed lines  other EoS's with this second PT of different length for the coexistence region could be considered as well.

\begin{figure}
     \centering
     \includegraphics[width=0.49\textwidth]{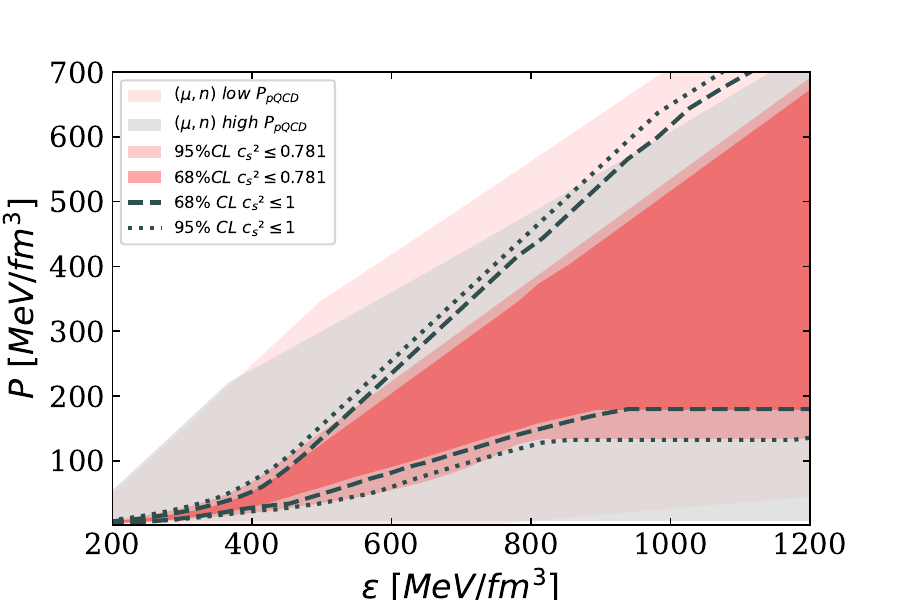} 
\includegraphics[width=0.49\textwidth]{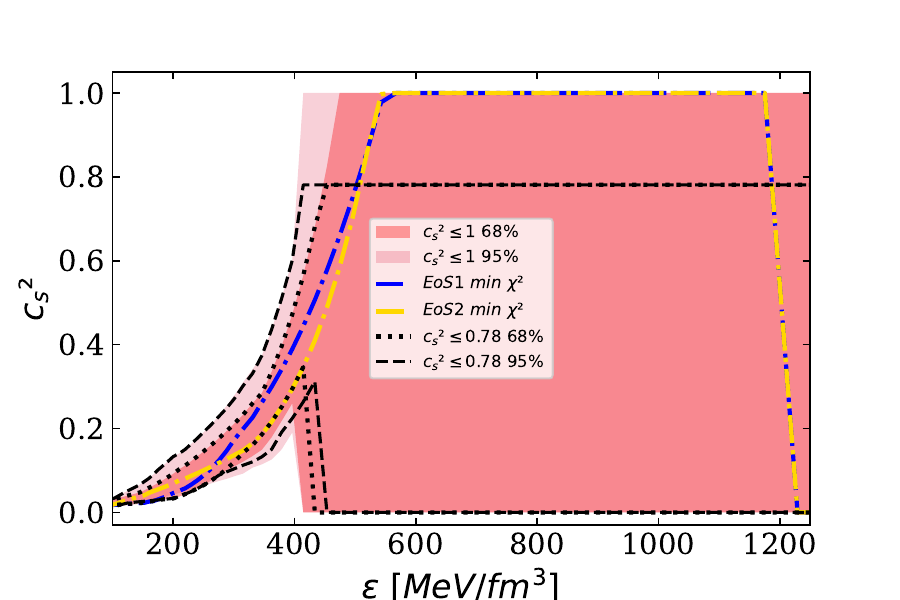} 
\caption{{\small All the EoS's considered in this plot are constrained by  astrophysical observables. {\bf Top:} Non-logarithmic bands of $(\varepsilon,p)$ up to typical central densities inside a NS. The red band (dotted line)  corresponds to our calculation with $c_s^2\leq 0.781(1.0)$ at the 68\% CL, and the pink one (dashed line) is similarly calculated at the  95\% CL. {\bf Bottom:} The distribution of $c_s^2$ as a
function of $\varepsilon$. The red and pink bands correspond to the calculation at the 68\% and 95\% CL for $c_s^2\leq 1$, respectively. In turn, the dashed and dotted lines correspond to $c_s^2\leq 0.781$ with the same statistical meaning. We also show by the dot-dashed lines the results for two EoS's with the smallest $\chi^2$ for $c_s^2\leq 1$. 
    \label{fig:comparebands} } }
 \end{figure}

The band of allowed EoS's for limiting $c_s^2 \leq 0.781$ \cite{Tang:2024jvs,Tang:2024jvs} is shown in the top panel of  Fig.~\ref{fig:comparebands} up to  typical central values of $\varepsilon$ in a NS. We observe that  until  $\varepsilon\approx 450$~MeV/fm$^3$ the resulting band is the same as the determined one in Fig.~\ref{fig:completeband}, but  above  this value the band starts  becoming  narrower. Now, for maximal central values we have  a chemical potential $ 1390\leq \mu_{\rm c,max} \leq 2020$~MeV at 68$\%$~CL and $ 1400\leq \mu_{\rm c,max} \leq 2030$~MeV at 95$\%$~CL, a baryon number density $5.2 n_s \leq n_{c} \leq 7.2 n_s$ at 68$\%$ CL and $4.75 n_s \leq n_{c} \leq 7.4 n_s$ at 95$\%$~CL, with $(E/A)_{\rm c,max} \approx 280-385~$MeV in both cases.    In this band of EoS we can find PTs with  starting densities between $2.51 n_s \leq n \leq n_{c}$. Within the interval of densities $2.51 n_s<n< 3.3 n_s$ at 68$\%$ CL, and between $2.51 n_s<n< 3.3 n_s$ at 95$\%$ CL, the PTS found show a coexistence range of baryon densities $ 0.15\leq \Delta n \leq 0.65 n_s$ of the short PT type.  In the case of long PTs, occurring within the density interval $3.3n_s\leq n\leq 13.5n_s$ and $3.2n_s\leq n\leq 14 n_s$  at 68$\%$ and 95$\%$, respectively, the coexistence region extended over $1 \lesssim \Delta n/n\leq 2.6$ at 68$\%$ CL, and $1  \lesssim \Delta n/n\leq 3$ at 95$\%$ CL. Thus, since  all these PTs report $1 \leq \Delta n/n$ they can be considered as strong first-order PTs \cite{Brandes:2023bob}. 
The maximum-mass difference obtained  between EoS's with limiting sound speed squared $c_s^2\leq 1$ and $c_s^2\leq 0.781$ is about 5--6$\%$, being smaller for the latter. Since the maximum limit for $c_s^2$ applies for large enough density energy values above around 420 and 520 MeV/fm$^3$ (which correspond to a star mass about $1.4~M_\odot$), the differences in the resulting values for the star radius can only be seen in the scale of the middle panel in Fig.~\ref{fig:completeband}  for $M>2.1$ $M_\odot$. For instance, at the 68\% CL, for $M=2.18\,M_\odot$ the maximum relative  difference in the radius found is $\lesssim 1\%$, and for $M=2.35\,M_\odot$ one has a maximal relative difference of  around $5.6\%$.

In the bottom panel of Fig.~\ref{fig:comparebands} we show the possible values of $c_s^2$ as a function of $\varepsilon$. The colored red and pink bands refer to the straight causal limit $c_s^2\leq 1$ distinguishing between the $1$ $\sigma$ and $2$ $\sigma$ regions, respectively. Similarly, the dashed and dotted lines enclose the $1$ and 2 $\sigma$ regions for the limit $c_s^2\leq 0.781$, in this order. The dot-dashed lines show the results for two  EoS's with the minimum $\chi^2$ found for $c_s^2\leq 1$. We observe that for $\varepsilon\lesssim 500$~MeV/fm$^3$ the sound velocity squared increases within a relatively narrow band, but once PTs are considered for $n>2.5 n_s$ all feasible  values for $c_s^2$ are registered by the resulting EoS. After the long PT for chemical potentials somewhat below $\mu_0<\mu_c$ the sound velocity squared is limited by the conformal value $c_s^2\leq 1/3$ to gradually reach the pQCD region.

\section{Summary and Conclusions}

In this work we have studied the equation of state (EoS) of a neutron star (NS) based on new results. On the theoretical side we use several recent developments of interest. In the low density range, for $n\leq 0.1 n_s$, we have the recent calculation of the pure neutron matter EoS at low densities obtained by Ref.~\cite{Alarcon:2022vtn}, being regulator independent and expressed directly in terms of experimental nucleon-nucleon scattering data. In the highest-density domain calculations in pQCD are now available up to ${\cal O}(\alpha_s^3)$ \cite{Kurkela:2009gj,Gorda:2021znl,Gorda:2021kme}. We then match between these two extremes by invoking first principles of  causality and  thermodynamic consistency and stability, following Refs.~\cite{LopeOter:2019pcq,Komoltsev:2021jzg}. On the experimental side, we take into account new measurements from PREX-II and CREX on the symmetry energy ($S_0$) and its slope ($L$) at saturation density, and observations and experimental measurements of masses and radii of different neutron stars.  
By enforcing the EoS to satisfy all these theoretical and experimental requirements we obtain a band of EoS's that describes the data of NSs and tidal deformabilities.  
The band of EoS's obtained also constrains further the possible values of $S_0$ and  $L$, narrowing the range of values of the latter by more than $50\%$. Our resulting values are  $32.9\leq S_0 \leq 39.5~\text{MeV}$ and $ 37.3 \leq L\leq 69.0~\text{MeV}$ at the 68\% CL. 
The band of EoS's constructed in this way also indicates  possible PTs for NS masses above 2.1~$M_\odot$ at 68\% CL, involving starting number densities above 2.5$n_s$. We find both long and short coexistence regions during the PTs, corresponding to first and second order ones, respectively.  Needless to say that the band of EoS's for NSs so determined  can be used for research related to study other NS properties and dark matter capture in NSs. 

We also calculate the band of EoS's when excluding the astrophysical observables in its determination,  constrained only by causality, stability, pQCD and nuclear physics. Interestingly, our results  can then be used to test GR and modified theories of gravity.
 
Our determined EoS's will be available at the CompOSE database (\url{https://compose.obspm.fr/manual/}) \cite{CompOSECoreTeam:2022ddl,Oertel:2016bki,Typel:2013rza}.

\section*{Acknowledgements}
We would like to thank Ingo~Tews for providing us the data to plot the green band in Fig.~\ref{fig.240613.1}. We also acknowledge partial financial support to the Grant PID2022-136510NB-C32 funded
by MCIN/AEI/10.13039/501100011033/ and FEDER, UE, and to the EU Horizon 2020 research and innovation program, STRONG-2020 project, under grant agreement no. 824093.

\bibliographystyle{apsrev4-1}
\bibliography{biblio}

\end{document}